%% file: hnl-2015-paper.tex
\documentclass[11pt]{article}
\usepackage{footnote}
\usepackage[dvips]{graphicx}
\usepackage{epsfig}
\usepackage{amssymb}
\usepackage{color}
\usepackage{xspace}
\usepackage{cite}
\usepackage[left]{lineno}

\def\geant  {\mbox{\textsc{Geant4}}\xspace}

\begin{document}
\centerline{\LARGE EUROPEAN ORGANIZATION FOR NUCLEAR RESEARCH}
%
%
\vspace{10mm} {\flushright{
CERN-EP-2017-311 \\
1 December 2017\\
\vspace{4mm}
Revised version:\\12 January 2018\\
}}
\vspace{-30mm}

%
%

%
\vspace{40mm}

\begin{center}
\boldmath
{\bf {\Large \boldmath{Search for heavy neutral lepton production in $K^+$ decays}}}
\unboldmath
\end{center}
\begin{center}
{\Large The NA62 Collaboration}\\
\end{center}

\begin{abstract}
A search for heavy neutral lepton production in $K^+$ decays using a data sample collected with a minimum bias trigger by the NA62 experiment at CERN in 2015 is reported. Upper limits at the $10^{-7}$ to $10^{-6}$ level are established on the elements of the extended neutrino mixing matrix $|U_{e4}|^2$ and $|U_{\mu4}|^2$ for heavy neutral lepton mass in the ranges 170--448~MeV/$c^2$ and 250--373~MeV/$c^2$, respectively. This improves on the previous limits from HNL production searches over the whole mass range considered for $|U_{e4}|^2$, and above 300~MeV/$c^2$ for $|U_{\mu4}|^2$.
\end{abstract}

\begin{center}
{\it Accepted for publication in Physics Letters B}
\end{center}

\newpage
\input{hnl2015_authors_6}
\newpage


\section*{Introduction}

Non-zero masses and mixing of the Standard Model (SM) neutrinos are now firmly established. However many SM extensions have been proposed, involving massive ``sterile'' neutrinos, also called heavy neutral leptons (HNLs), which mix with the ordinary light ``active'' neutrinos. For example, the Neutrino Minimal Standard Model ($\nu$MSM) postulates three HNLs, explaining dark matter and baryon asymmetry of the universe in a way consistent with the results of neutrino oscillation experiments~\cite{nuMSM}. One of these HNLs with the expected mass of ${\cal O}(10~{\rm keV}/c^2)$ is a dark matter candidate, while the others are expected to have masses of ${\cal O}(1~{\rm GeV}/c^2)$.

Mixing between HNLs (denoted $N$ below) and active light neutrinos gives rise to HNL production in meson decays, including $K^+\to\ell^+N$ ($\ell=e,\mu$). The branching fraction of the latter decay is determined by the HNL mass $m_N$ and mixing parameter $|U_{\ell 4}|^2$ as follows~\cite{sh80,sh81}:
\begin{equation}
{\cal B}(K^+\to\ell^+ N) = {\cal B}(K^+\to\ell^+\nu) \cdot \rho_\ell(m_N) \cdot |U_{\ell 4}|^2.
\label{eq:main}
\end{equation}
Here ${\cal B}(K^+\to\ell^+\nu)$ is the measured branching fraction of the SM leptonic decay (including inner bremsstrahlung), and $\rho_\ell(m_N)$ is a kinematic factor:
\begin{displaymath}
\rho_\ell(m_N) = \frac {(x+y)-(x-y)^2} {x(1-x)^2} \cdot \lambda^{1/2}(1,x,y),
\end{displaymath}
with $x=(m_\ell/m_K)^2$, $y=(m_N/m_K)^2$ and $\lambda(a,b,c)=a^2+b^2+c^2-2(ab+bc+ac)$. By definition, $\rho_\ell(0)=1$. Numerically, the product ${\cal B}(K^+\to\ell^+\nu) \cdot \rho_\ell(m_N)$ is ${\cal O}(1)$ over most of the allowed $m_N$ range. However it drops to zero at the kinematic limit $m_N=m_K-m_\ell$ and, in the positron case, reduces to ${\cal B}(K^+\to e^+\nu)=1.582(7)\times 10^{-5}$~\cite{pdg} for $m_N \to 0$ due to helicity suppression.


A search for $K^+\to\ell^+N$ decays in HNL mass range 170--448~MeV/$c^2$ using a data sample collected with a minimum bias trigger by the NA62 experiment at CERN during the first physics data-taking in 2015 is reported here. The obtained upper limits on $|U_{\ell4}|^2$ complement, and improve on, those obtained in earlier HNL production searches in pion and kaon decays~\cite{br92,ag17,ya84,ar15,la17}.

\section{Beam, detector and data sample}
\label{sec:detector}

\begin{figure}[t]
\begin{center}
\resizebox{\textwidth}{!}{\includegraphics{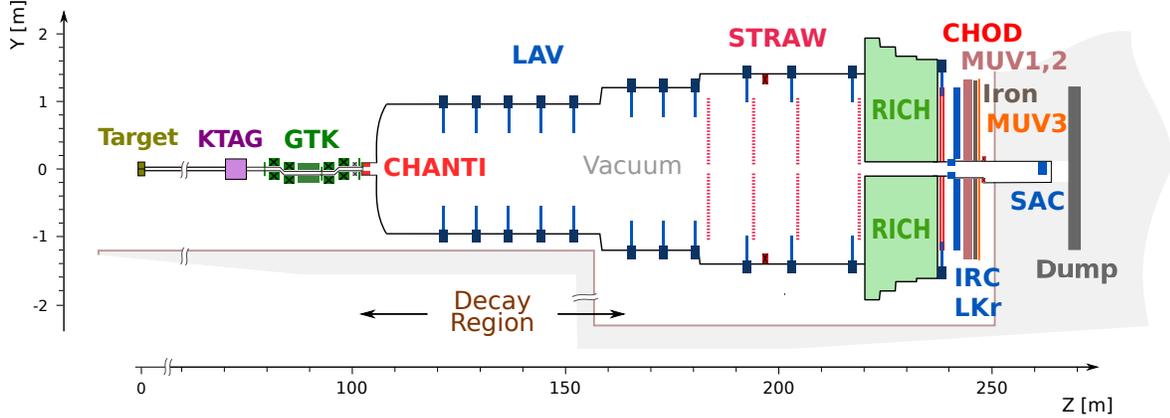}}%
\end{center}
\vspace{-13mm}
\caption{Schematic side view of the NA62 beamline and detector.}
\label{fig:detector}
\end{figure}

The layout of the NA62 beamline and detector~\cite{na62-detector} is shown schematically in Fig.~\ref{fig:detector}. A secondary positive hadron beam with a central momentum of 75 GeV/$c$ and 1\% momentum spread (rms) is derived from primary 400 GeV/$c$ protons extracted from the CERN SPS and interacting with a beryllium target in spills of 3~s effective duration at nominal intensity of $1.1\times 10^{12}$~protons/s. Beam kaons are tagged with a 70~ps time resolution by a differential Cherenkov counter (KTAG) with nitrogen radiator at 1.73~bar pressure contained in a 5~m long vessel. Beam particle momenta are measured by a silicon pixel detector (GTK, under commissioning in 2015 and not used for this analysis). Inelastic interactions of beam particles with the last of the three GTK stations are detected by an array of scintillator hodoscopes (CHANTI). The beam is delivered into a vacuum tank containing a 75~m long fiducial decay volume (FV) starting 2.6~m downstream of the last GTK station. The beam transverse size at the FV entrance is $53\times24~{\rm mm}^2$, and the beam divergence in 2015 was 0.22~(0.11)~mrad in the horizontal (vertical) plane. The nominal instantaneous particle rate at the FV entrance is 750~MHz, mainly due to $\pi^+$ (70\%), protons (23\%) and $K^+$ (6\%). The fraction of kaons decaying in the FV is $13\%$, leading to 6~MHz nominal $K^+$ decay rate. The beam is accompanied by a flux of muons produced by $K^+$ and $\pi^+$ decays upstream of the vacuum tank (the beam halo), with 3~MHz nominal rate in the detector acceptance. Central holes in detectors downstream of the FV and a beam pipe traversing most of these detectors allow the undecayed beam particles to continue their path in vacuum.
The beam intensity during the 2015 run was typically ${\cal O}(1\%)$ of the nominal value.

The momenta of charged $K^+$ decay products are measured by a magnetic spectrometer (STRAW) located in the vacuum tank downstream of the FV. The spectrometer consists of four tracking chambers made of straw tubes, and a dipole magnet located between the second and the third chamber providing a horizontal momentum kick of approximately $270~\mathrm{MeV}/c$. The spectrometer momentum resolution is $\sigma_p/p = (0.30\oplus 0.005\cdot p)\%$, where the momentum $p$ is expressed in GeV/$c$.

A $27X_0$ thick quasi-homogeneous liquid krypton (LKr) electromagnetic calorimeter, built for the earlier NA48 experiment~\cite{na48-detector} and equipped with a new readout system, is used for photon detection. The calorimeter has an active volume of 7~m$^3$, and is segmented transversally into 13248 projective $\sim\!2\!\times\!2$~cm$^2$ cells. Its energy resolution in the NA62 conditions is $\sigma_E/E=(4.8/\sqrt{E}\oplus11/E\oplus0.9)\%$, where $E$ is expressed in GeV. To achieve hermetic acceptance for photons emitted in $K^+$ decays in the FV at angles up to 50~mrad, the LKr calorimeter is supplemented by annular lead glass large-angle veto (LAV) detectors installed in 12~positions along and downstream of the FV, and two lead/scintillator sampling calorimeters (intermediate-ring calorimeter, IRC, and small-angle calorimeter, SAC) located close to the beam axis.

A ring-imaging Cherenkov detector (RICH) consisting of a 17.5~m long vessel filled with neon at atmospheric pressure is used for identification of charged $K^+$ decay products. Its Cherenkov threshold for muons is 9.5~GeV/$c$, and it provides timing measurement for tracks above threshold to better than 100~ps precision. The LKr calorimeter, a hadronic iron/scintillator sampling calorimeter formed of two modules (MUV1,2) and a scintillator-tile muon detector (MUV3) located behind an 80 cm thick iron wall are also used for particle identification. A plastic scintillator hodoscope (CHOD) built for the NA48 experiment, located in front of the calorimeters, provides a fast trigger with efficiency above 99\% and track-timing measurement to 200~ps precision.

The data sample used for this analysis is obtained from $1.2\times 10^4$ SPS spills recorded in 5~days of operation in 2015 at beam intensity varying from 0.4\% to 1.3\% of the nominal value with a minimum-bias trigger scheme. The low-level hardware trigger required a CHOD signal (downscaled typically by a factor of 3) to collect $K^+$ decays to muons (which account for 67\% of the decay rate), and a CHOD signal in anti-coincidence with a MUV3 signal (not downscaled) to collect decays with no muons in the final state. The high-level software trigger required a kaon signal in the KTAG detector within $\pm10$~ns of the low-level trigger signal. Loose timing conditions are used in this analysis because the accidental rates are small, due to the low beam intensity.

\section{Event selection}
\label{sec:selection}


Assuming $|U_{\ell4}|^2<10^{-4}$ and considering HNL decays into SM particles~\cite{go07}, the smallest possible average decay length of a HNL produced in the $K^+\to\ell^+N$ decays in NA62 conditions exceeds 10~km. Under the above assumption, HNL decays in flight in the 156~m long volume from the start of the FV to the last detector (SAC) can be neglected, and the $K^+\to\ell^+N$ decay is characterized by a single detected track in the final state, similarly to the SM $K^+\to\ell^+\nu$ decay. The principal selection criteria are listed below.
%
%
\begin{itemize}
\item A single positively charged track reconstructed in the spectrometer with momentum in the range 5--70~GeV/$c$ is required. Additional spectrometer tracks and LKr energy deposition clusters not geometrically compatible with the track are not allowed within $\pm 100$~ns of the track time measured by the CHOD. Activity in the large-angle and small-angle photon veto detectors and the CHANTI detector within $\pm 10$~ns of the track time is not allowed. Track impact points in the straw chambers, LKr calorimeter, CHOD and MUV1--3 detectors should be within their fiducial geometrical acceptances.
\item The kaon decay vertex is reconstructed as the point of closest approach of the track and the beam axis (the latter is monitored with fully reconstructed $K^+\to\pi^+\pi^+\pi^-$ decays), taking into account the measured stray magnetic field map in the vacuum tank. The reconstructed closest distance of approach (CDA) between the track and beam axis should be less than 25~mm, as determined by the beam transverse size.
\item To suppress beam halo background from $K^+$ decays upstream of the KTAG and beam $\pi^+$ decays, the presence of a kaon signal in the KTAG is required within $\pm 10$~ns of the track time measured by the CHOD.
\item Beam halo background from $K^+\to\mu^+\nu$ decays over the approximately 30~m long path between the KTAG and the last GTK station (with the muon deflected by  magnetic fields and scattered in magnet yokes and collimators before reaching the vacuum tank) is suppressed by geometrical conditions established by studies of upstream $K^+$ decays. For the $K^+\to e^+N$ selection, the reconstructed vertex position is required to be at least 10~m downstream of the start of the FV. For the $K^+\to\mu^+N$ selection, the minimal required distance between the decay vertex and the start of the FV depends on the muon emission angle with respect to the beam axis and lies in the range 10--33~m.
\item
Positrons and muons are identified by the ratio of energy deposit, $E$, in the LKr calorimeter to momentum, $p$, measured by the spectrometer: \mbox{$0.9<E/p<1.15$} and $E/p<0.2$, respectively. No signals in MUV1--3 detectors within $\pm20$~ns of the track time and geometrically consistent with $e^+$ candidate tracks (accounting for detector granularity and multiple scattering) are allowed; MUV1--3 signals are required for $\mu^+$ candidate tracks. Additionally, an identification algorithm based on the RICH hit pattern is applied for tracks with \mbox{$p<40$~GeV/$c$}.
\end{itemize}

The squared missing mass is computed as $m_{\rm miss}^2=(P_K-P_\ell)^2$, where $P_K$ and $P_\ell$ are the kaon and lepton 4-momenta, respectively. $P_K$ is obtained from the beam average 3-momentum (monitored with $K^+\to\pi^+\pi^+\pi^-$ decays) in the $K^+$ mass hypothesis, while $P_\ell$ is evaluated from the reconstructed track 3-momentum in the corresponding $\ell^+$ mass hypothesis.

\begin{figure}[t]
\begin{center}
\resizebox{0.50\textwidth}{!}{\includegraphics{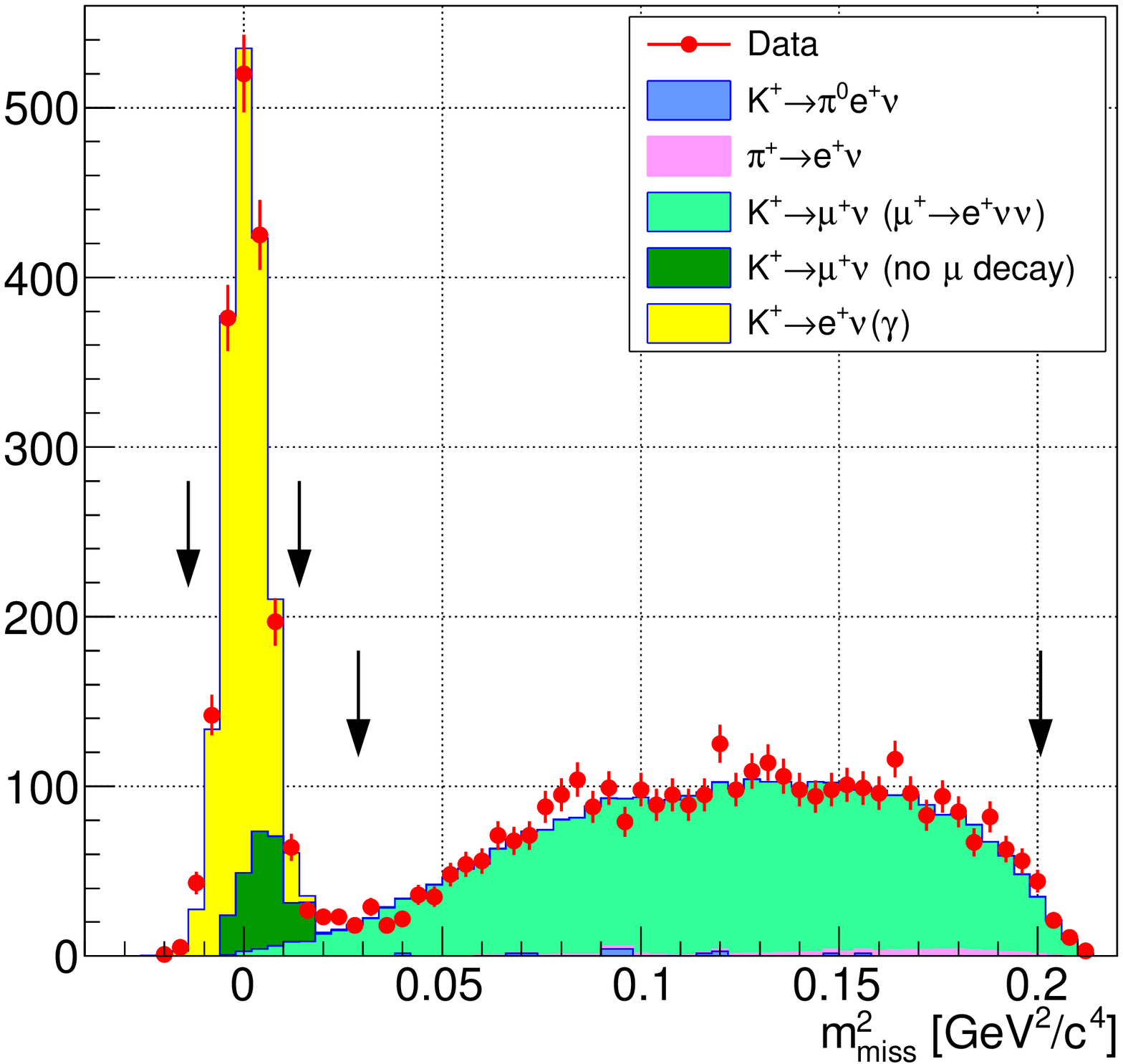}}%
\resizebox{0.50\textwidth}{!}{\includegraphics{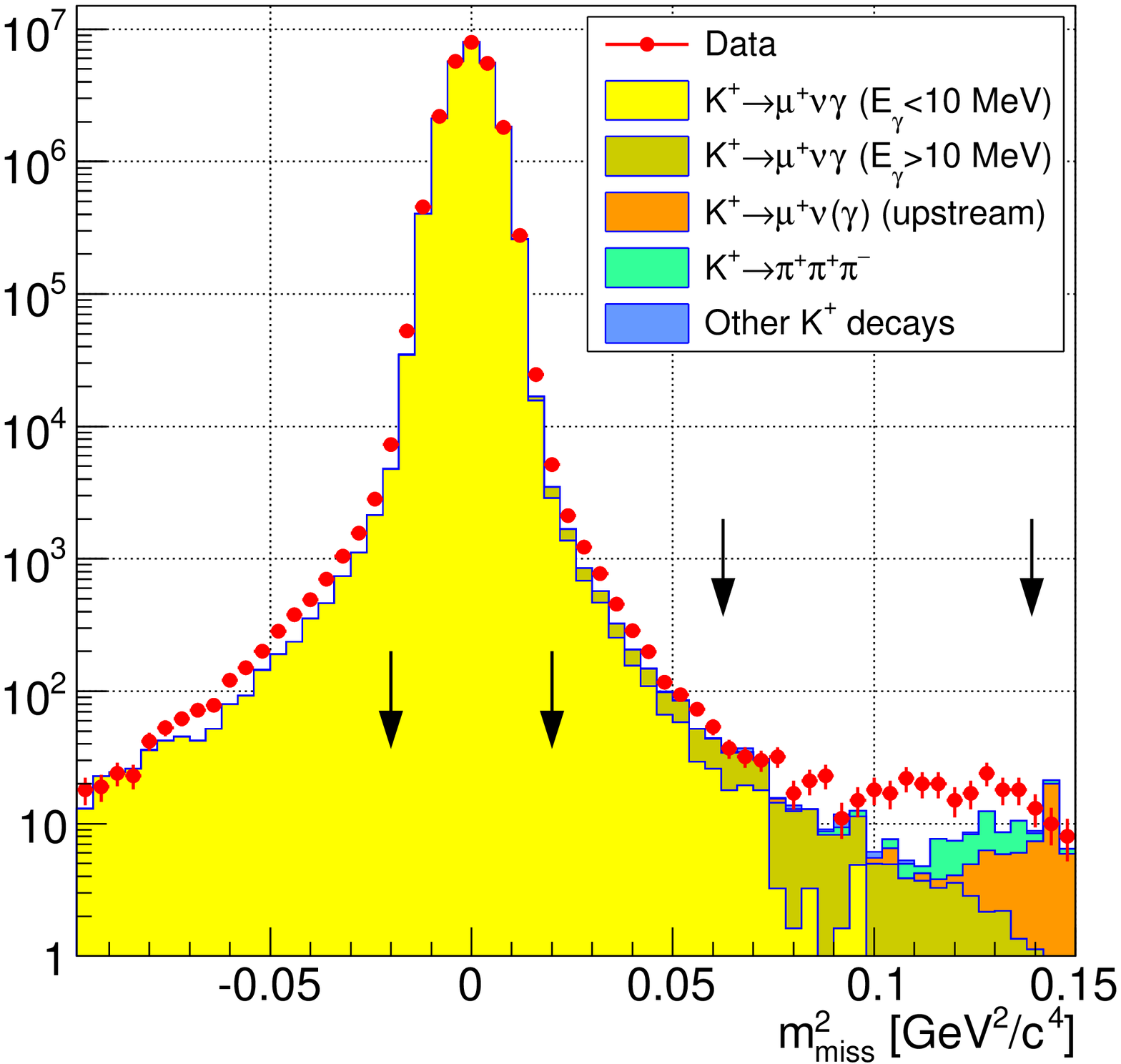}}
\put(-432,200){\bf\large (a)} \put(-204,200){\bf\large (b)}
\end{center}
\vspace{-13mm}
\caption{Distributions of the $m_{\rm miss}^2$ variable for data and simulated events passing the (a) $e^+$ and (b) $\mu^+$ selections. The bin widths are $0.004~{\rm GeV}^2/c^4$. Pairs of vertical lines in each plot indicate the boundaries of the SM and HNL signal regions. The HNL signal regions defined in Section~\ref{sec:selection} correspond approximately to 0.03--0.20~${\rm GeV}^2/c^4$ and 0.06--0.14~${\rm GeV}^2/c^4$ in $m_{\rm miss}^2$ values in the $e^+$ and $\mu^+$ case, respectively.}
\label{fig:mmiss2}
\end{figure}

Simulation of particle interactions with the detector and its response is performed with a Monte Carlo (MC) simulation package based on the \geant toolkit~\cite{geant4}. The $m_{\rm miss}^2$ spectra of the selected events from both data and simulation are displayed in Fig.~\ref{fig:mmiss2}. Signals from the SM leptonic decays \mbox{$K^+\to\ell^+\nu$} are observed as peaks at $m_{\rm miss}^2=0$ with $m_{\rm miss}^2$ resolutions of $4.9~(4.7)\times 10^{-3}~{\rm GeV}^2/c^4$ in the $e^+$ ($\mu^+$) case. These resolutions are dominated by the  momentum spread and divergence of the beam, and are reproduced by MC simulations to 1\% relative precision. The SM and HNL signal regions are defined in the $e^+$ ($\mu^+$) case as \mbox{$|m_{\rm miss}^2|<0.014~(0.020)~{\rm GeV}^2/c^4$} and \mbox{$170~(250)<m_{\rm miss}<448~(373)$~MeV/$c^2$}, respectively. The search for $K^+\to\ell^+N$ decays consists of a search for peaks above background in the HNL signal regions.

\section{Measurement principle}
\label{sec:flux}

The $K^+\to\ell^+N$ decay rates are measured with respect to the rates of the normalization SM $K^+\to\ell^+\nu$ decays with similar topologies and known branching fractions. The expected numbers of $K^+\to\ell^+N$ signal events $N_S^\ell$ are related to the assumed branching fractions ${\cal B}(K^+\to\ell^+N)$ and acceptances $A_\ell^N$ of the $K^+\to\ell^+N$ selections as
\begin{equation}
N_S^\ell= N_K^\ell \cdot {\cal B}(K^+\to\ell^+N) \cdot A_\ell^N.
\label{eq:master}
\end{equation}
Here $N_K^\ell$ are the numbers of $K^+$ decays in the FV, computed from the numbers $N_\ell$ of selected data events with $m_{\rm miss}^2$ in the SM signal region:
\begin{displaymath}
N_K^e = \frac{N_e}{ A_e^e \cdot {\cal B}(K^+\to e^+\nu) + A_e^\mu \cdot {\cal B}(K^+\to\mu^+\nu)} = (3.00\pm0.11)\times 10^8
\end{displaymath}
and
\begin{displaymath}
N_K^\mu = \frac{N_\mu}{A_\mu^\mu \cdot {\cal B}(K^+\to\mu^+\nu)} = (1.06\pm0.02)\times 10^8,
\end{displaymath}
where $A_{\ell_1}^{\ell_2}$ is the acceptance of the $K^+\to\ell_1^+\nu$ selection (with $m_{\rm miss}^2$ in the SM signal region) for the $K^+\to\ell_2^+\nu$ decay evaluated with MC simulations, and ${\cal B}(K^+\to\ell^+\nu)$ is the branching fraction of the $K^+\to\ell^+\nu$ decay~\cite{pdg}. The inputs to the computation of $N_K^\ell$ are summarized in Table~\ref{tab:flux}. The number of $K^+$ decays in the $\mu^+$ case is smaller than that in the $e^+$ case due to the downscaling factor of typically 3 applied to the muon trigger chain.

The above approach relies on first-order cancellation between signal, normalization and background yields of the effects of residual detector inefficiencies, trigger efficiencies and random veto not fully accounted for by the MC simulation.


\begin{table}[tb]
\caption{Inputs to the computation of the numbers $N_K^\ell$ of kaon decays in the FV: numbers of selected data events in the SM signal region, acceptances evaluated with MC simulations and their statistical errors (notation is defined in the text), and $K^+\to\ell^+\nu$ branching fractions~\cite{pdg}.}
\begin{center}
\vspace{-9mm}
\begin{tabular}{lcc}
\hline
& $K^+\to e^+\nu$ selection & $K^+\to\mu^+\nu$ selection\\
\hline
Number of data events $N_\ell$ & 1767 & $2.403\times 10^{7}$\\
Acceptance $A^\mu_\ell$ & $(1.30\pm0.17)\times 10^{-6}$ & $0.3579\pm0.0001$ \\
Acceptance $A^e_\ell$ & $0.3197\pm0.0008$ & -- \\
${\cal B}(K^+\to\ell^+\nu)$ & $(1.582\pm0.007)\times 10^{-5}$ & $0.6356\pm0.0011$ \\
\hline
\end{tabular}
\end{center}
\vspace{-12mm}
\label{tab:flux}
\end{table}

The background in the $K^+\to e^+\nu$ sample from $K^+\to\mu^+\nu$ decays, due to both $\mu^+$ mis-identification and decay in flight, is taken into account in the computation of $N_K^e$. This background is dominated by $\mu^+$ mis-identification due to `catastrophic' bremsstrahlung in the LKr calorimeter at track momenta $p>40~{\rm GeV}/c$, where identification relies on calorimetry only as the RICH does not provide useful information. The probability of a muon having $E/p>0.90$ in the LKr calorimeter has been measured in a dedicated study to be ${\cal P}_{\mu e}\sim 10^{-5}$, and found to be reproduced by simulation to 10\% relative precision~\cite{la11}. The background in the $K^+\to\mu^+\nu$ sample is negligible.

The quoted uncertainty on $N_K^e$ receives contributions from the statistical error (2.4\%), precision on the simulation (evaluated by stability checks versus variation of the selection conditions and considering the precision on ${\cal P}_{\mu e}$ simulation, 2.0\%), MC statistical precision on the acceptance for the $K^+\to\mu^+\nu$ background (1.9\%) and the external parameter ${\cal B}(K^+\to e^+\nu)$ (0.4\%), combined in quadrature to obtain a total relative error of 3.7\%. The uncertainty on $N_K^\mu$ receives two contributions of similar size: due to the precision of the simulation (evaluated by variation of the selection conditions) and due to the external input ${\cal B}(K^+\to \mu^+\nu)$, combined in quadrature to obtain a total relative error of 1.9\%.


\section{Background estimates with MC simulations}
\label{sec:bkg}

The HNL search procedure, presented in Section~\ref{sec:search}, is based on a data-driven background estimation method, but this is only valid provided there are no peaking background structures in the HNL mass region. Backgrounds to HNL production have been estimated by MC simulations (Fig.~\ref{fig:mmiss2}) to understand qualitatively their contributions and to optimize the event selection. The results of these simulation studies, reported below, justify the adopted procedure.
%
%

\boldmath
\subsection{Backgrounds to $K^+\to e^+N$}
\label{sec:bkg-ke2}
\unboldmath

The principal background to $K^+\to e^+N$ decays comes from the $K^+\to\mu^+\nu$ decay followed by muon decay in flight $\mu^+\to e^+\nu\bar\nu$. It is characterized by a broader CDA distribution than the signal, and is suppressed by the CDA and vertex position selection criteria (Section~\ref{sec:selection}). The CDA selection criterion and therefore the background level are determined by the beam transverse size. The background due to $K^+\to\mu^+\nu$ decays with muon mis-identification (Section~\ref{sec:flux}) is constrained to low $m_{\rm miss}^2$ values outside the HNL signal region.

Beam pion decays $\pi^+\to e^+\nu$, as well as $\pi^+\to\mu^+\nu$ followed by muon decay in flight, contribute to the background via $\pi^+$ mis-identification by the KTAG due to accidental coincidence with a beam kaon not decaying in the FV. The contribution from direct $\pi^+$ mis-identification by the KTAG is negligible. Pion mis-identification probability for the employed KTAG--CHOD timing condition, averaged over the data sample, is computed to be $(0.9\pm0.1_{\rm syst})\%$ from the beam $K^+$ rate measured via the rate of out-of-time $K^+$ signals in the KTAG. This estimate is consistent with the number of observed $\pi^+\to e^+\nu$ decays in the $(P_\pi-P_e)^2$ spectrum, where $P_\pi$ is the beam pion 4-momentum.


Backgrounds from all other major $K^+$ decays with branching fractions above 1\%, and all $K^+$ decays to positrons and branching fractions above $10^{-5}$~\cite{pdg} have been considered. The $m_{\rm miss}^2$ spectra of the estimated background components are displayed in Fig.~\ref{fig:mmiss2}a, showing good agreement with the data spectrum.


\boldmath
\subsection{Backgrounds to $K^+\to\mu^+N$}
\label{sec:bkg-kmu2}
\unboldmath

The largest component of the background to $K^+\to\mu^+N$ decays comes from the  $K^+\to\mu^+\nu\gamma$ decay, mainly due to photons emitted at angles greater than 50~mrad with respect to the beam axis and escaping the LAV geometrical acceptance. It is simulated including inner bremsstrahlung and structure-dependent processes as well as their interference~\cite{bi92}; decays with the photon energy in the kaon rest frame $E_\gamma$ below and above 10~MeV are simulated separately to increase the MC statistics in the latter case.

Residual background due to $K^+$ decays between the KTAG and the last GTK station, which is suppressed by the cut on the vertex longitudinal position (Section~\ref{sec:selection}), is estimated from a dedicated simulation. Backgrounds from all other major $K^+$ decays are also considered: the largest of them is due to the $K^+\to\pi^+\pi^+\pi^-$ decay.

The $m_{\rm miss}^2$ spectra of the estimated background components are displayed in Fig.~\ref{fig:mmiss2}b; current agreement with the data spectrum in the HNL signal region is marginal. Observation in the data of a background component (not reproduced with MC simulation) with muons propagating close to the $yz$ plane (which is the bending plane of the GTK dipole magnets) suggests that the data/MC disagreement in the HNL signal region is due to the limited precision on the beamline simulation affecting the estimated background from upstream $K^+$ decays. On the other hand, the disagreement at negative $m_{\rm miss}^2$ is due to the limited precision of the description of the resolution, affected by the simulation of the beam momentum spectrum and divergence.


\begin{figure}[t]
\begin{center}
\resizebox{0.50\textwidth}{!}{\includegraphics{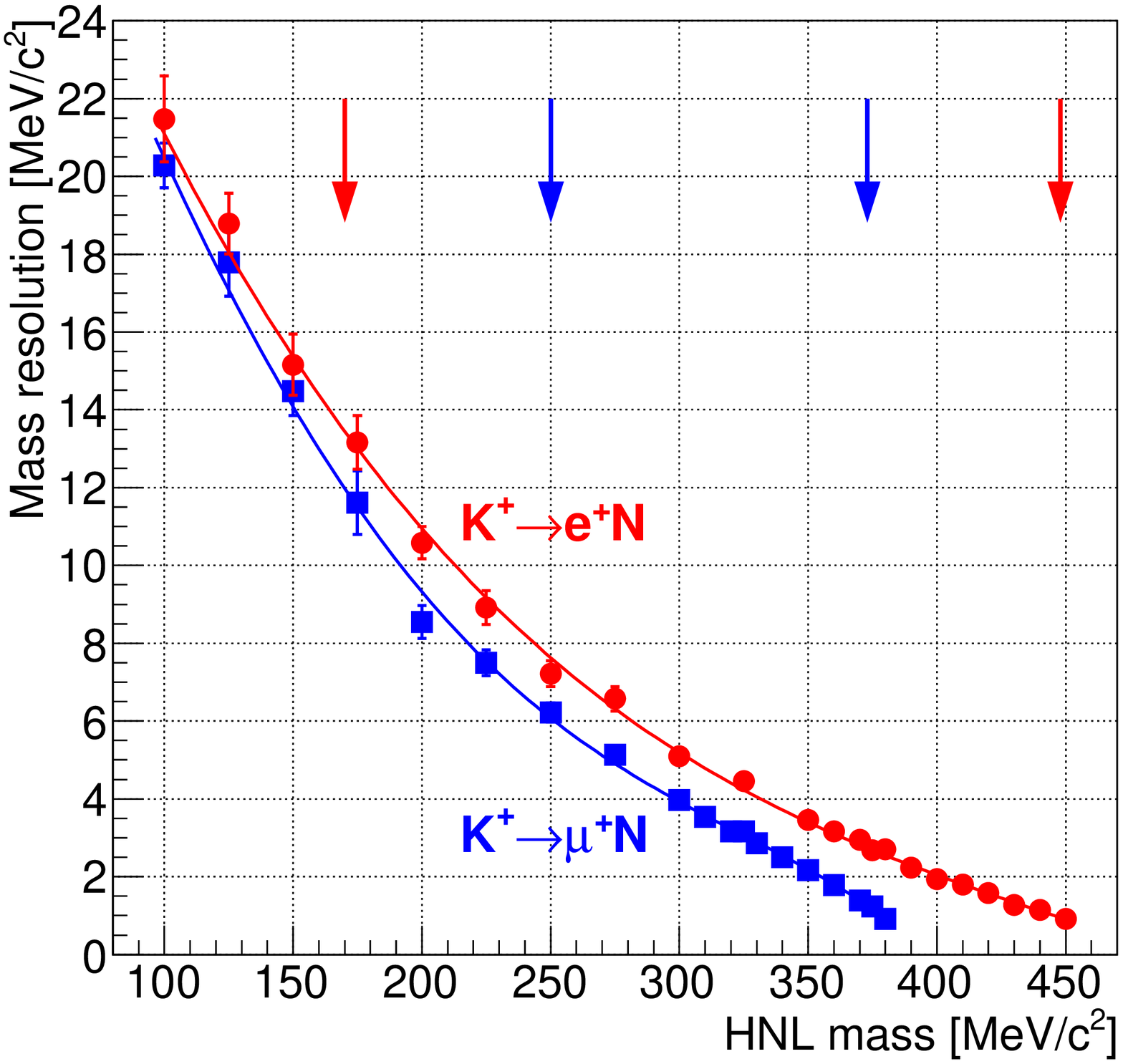}}%
\resizebox{0.50\textwidth}{!}{\includegraphics{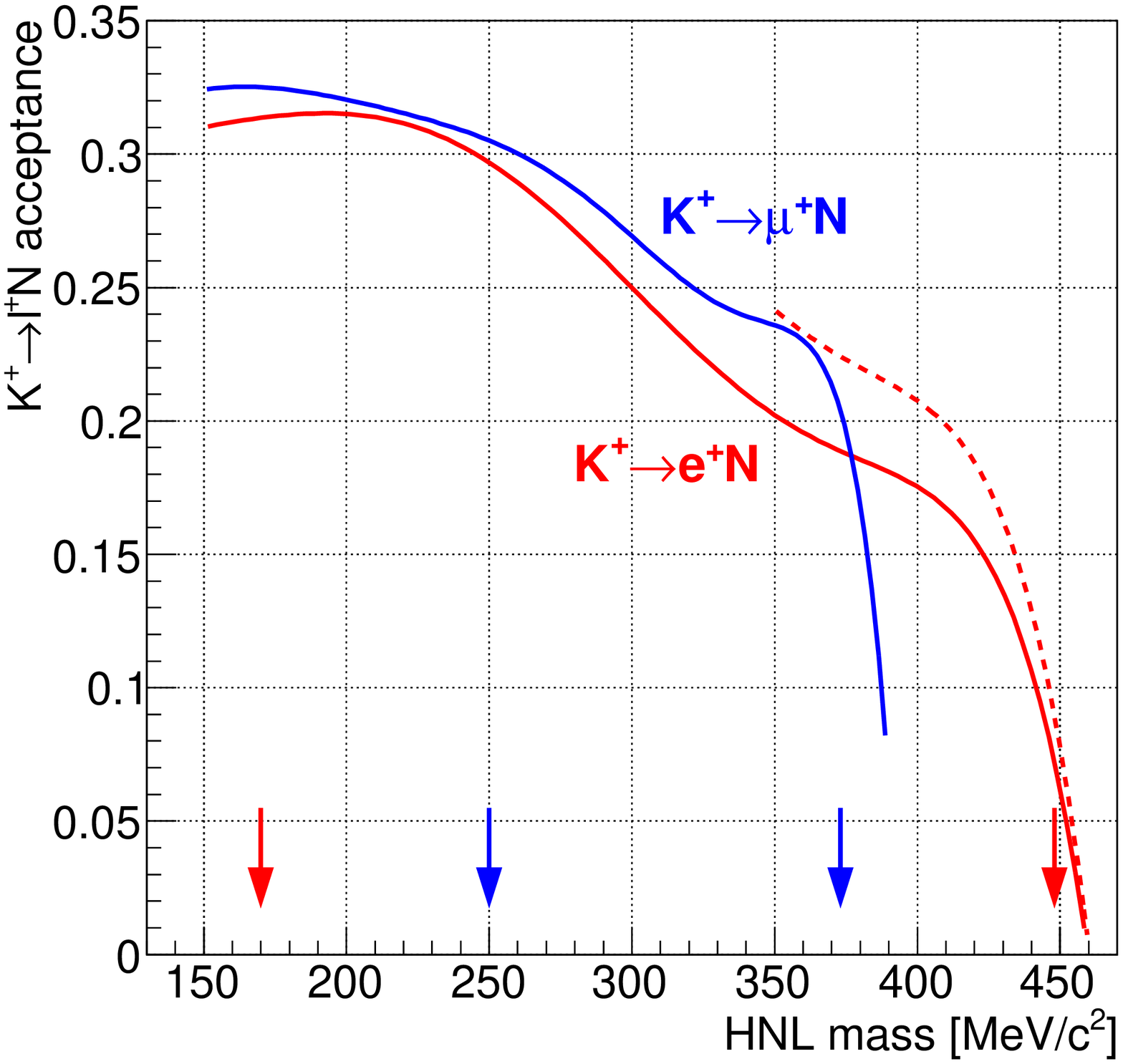}}
\put(-250,200){\bf\large (a)} \put(-21,200){\bf\large (b)}
\end{center}
\vspace{-16mm}
\caption{(a) Missing mass resolution $\sigma_m^\ell$ evaluated from MC simulations: values obtained for a set of HNL masses with their statistical errors, and polynomial functions used to define the HNL selection criterion. The corresponding resolution on the squared missing mass in the signal regions is a few $10^{-3}$~GeV$^2/c^4$, and has a weak mass dependence. (b) Acceptances $A_\ell^N$ of the $K^+\to\ell^+N$ selections obtained from MC simulations; the dashed line corresponds to the loose $K^+\to e^+N$ selection applied for HNL masses of 350~MeV/$c^2$ and above. Vertical arrows indicate the extent of the HNL signal regions.}
\vspace{-2mm}
\label{fig:resolution-acceptance}
\end{figure}

\section{Search for HNL production}
\label{sec:search}

Mass scans are performed in the HNL signal regions with a step size of 1~MeV/$c^2$. The event selection employed for each HNL mass hypothesis involves an additional condition: the reconstructed missing mass should be within $\pm1.5\sigma_m^\ell$ of the assumed HNL mass, where $\sigma_m^\ell$ is the mass resolution evaluated with MC simulations (Fig.~\ref{fig:resolution-acceptance}a). The above width of the signal mass window leads to near-optimal expected upper limits on ${\cal B}(K^+\to\ell^+N)$ in the absence of signals across the whole HNL signal regions. A loose selection with a relaxed vertex longitudinal position constraint (requiring the vertex to be in the FV) is applied in the $K^+\to e^+N$ case for mass hypotheses of 350~MeV/$c^2$ and higher, reflecting the fact that the beam halo background does not populate this mass range. Acceptances, $A_{\ell}^N$, of the selections (including the $\pm1.5\sigma_m^\ell$ mass cut) as functions of HNL mass obtained with MC simulations are shown in Fig.~\ref{fig:resolution-acceptance}b.

To certify that the missing mass resolution and therefore the signal acceptance are simulated correctly outside the SM $K^+\to\ell^+\nu$ peaks, the resolution on $\Delta m^2_{3\pi} = (P_K-P_3)^2-(P_1+P_2)^2$ in fully reconstructed $K^+\to\pi^+\pi^+\pi^-$ decays, where $P_{i}$ ($i=1,2,3$) are the pion 4-momenta reconstructed from the spectrometer information and $P_K$ is the kaon 4-momentum defined as for the $m_{\rm miss}^2$ computation (Section~\ref{sec:selection}), has been studied as a function of $(P_1+P_2)^2$. Given that $\Delta m^2_{3\pi}=0$ by construction, the resolution on $\Delta m^2_{3\pi}$ can be measured for both data and MC. Data and MC resolutions have been found to agree within 1\%. For the adopted $\pm1.5\sigma_m^\ell$ mass window, 1\% change on $\sigma_m^\ell$ translates into 0.4\% relative change on the signal acceptance.

In each HNL mass hypothesis considered, the background is evaluated from sidebands of the data $m_{\rm miss}$ distribution. The number of expected background events $N_{\rm exp}$ within the $\pm1.5\sigma_m^\ell$ HNL search window is estimated from a least-squares fit to the data $m_{\rm miss}$ spectrum with a bin width of 5~MeV/$c^2$ using third order polynomial functions in the 100--460~(200--385)~MeV/$c^2$ range for the $e^+$ ($\mu^+$) case. Mass bins overlapping with the $\pm1.5\sigma_m^\ell$ wide HNL search window are excluded from the fit to avoid bias caused by possible HNL signals. Statistical uncertainties $\delta N_{\rm exp}$ on the background estimates $N_{\rm exp}$ are computed by propagation of statistical errors on the fit function parameters: they are typically about 10\% in relative terms. Systematic uncertainties on $N_{\rm exp}$ due to the choice of background fit function, estimated by using fourth order polynomials for the fits, are negligible (typically 1\%).

\begin{figure}[p]
\begin{center}
\resizebox{0.50\textwidth}{!}{\includegraphics{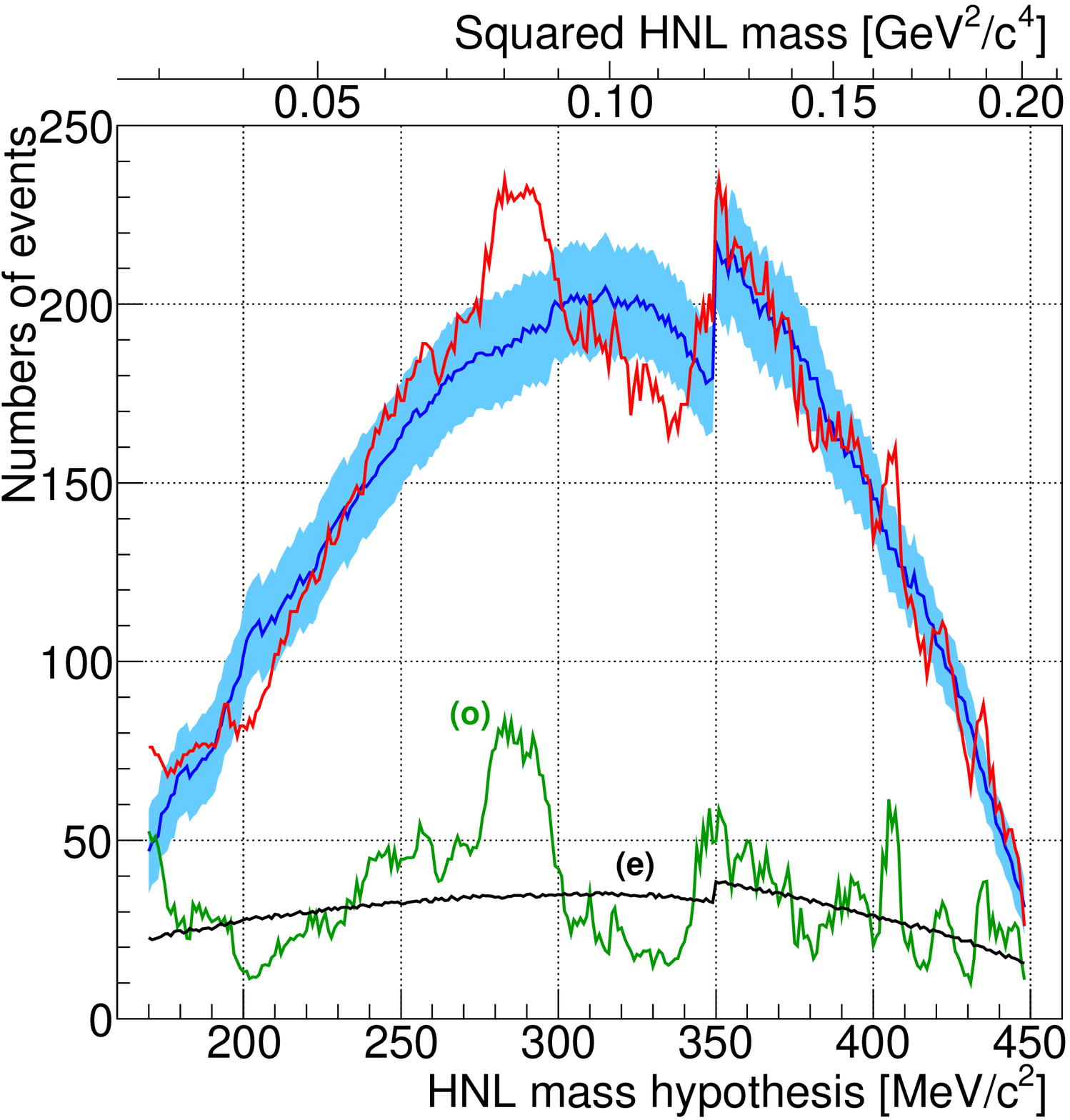}}%
\resizebox{0.50\textwidth}{!}{\includegraphics{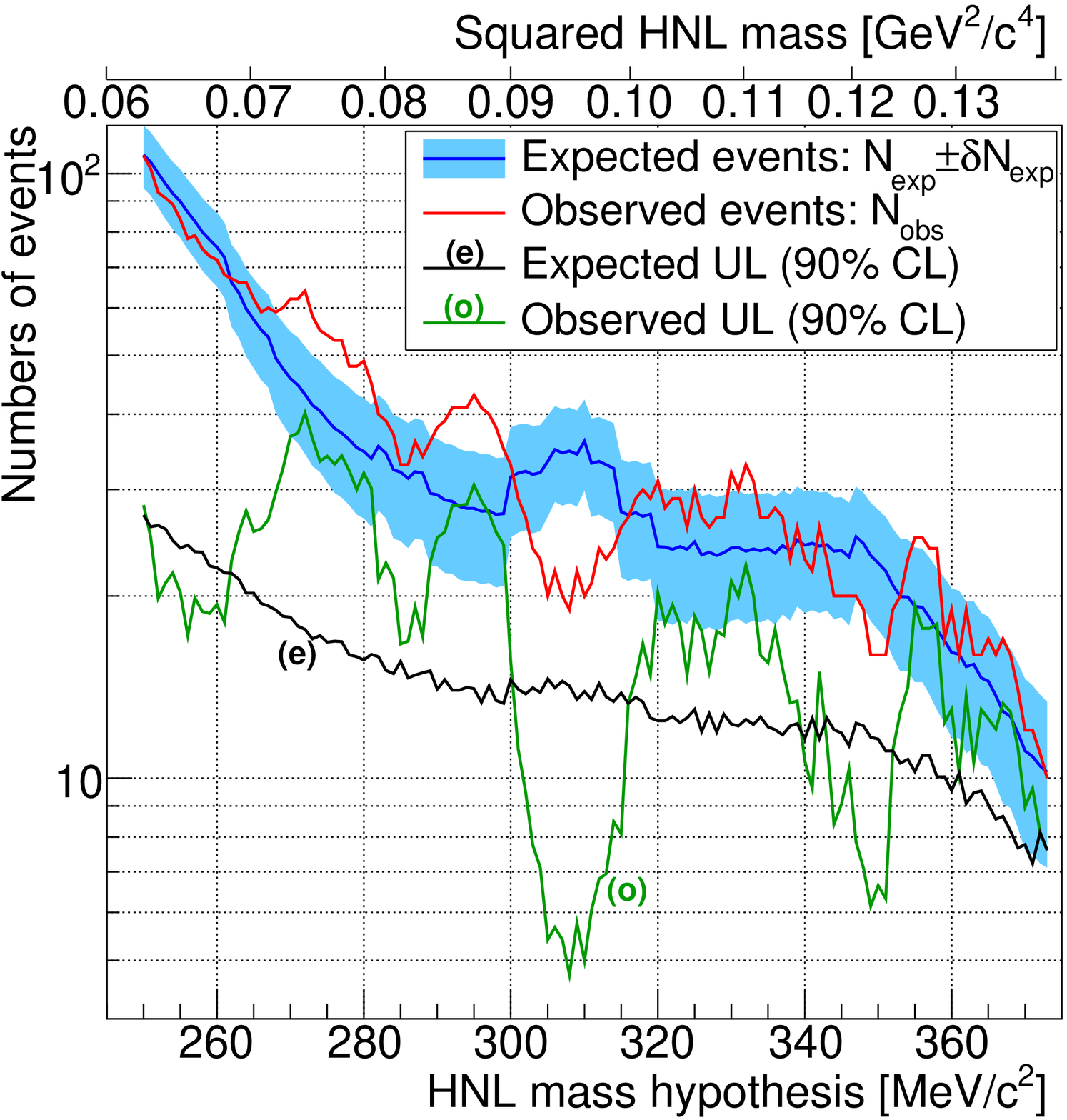}}
\put(-423,200){\bf\large (a)} \put(-180,200){\bf\large (b)}
\end{center}
\vspace{-12mm}
\caption{For each NHL mass hypothesis, numbers of expected ($N_{\rm exp}$) and observed ($N_{\rm obs}$) events, together with the uncertainty on $N_{\rm exp}$ ($\delta N_{\rm exp}$, as shown by the blue band); expected and observed upper limits at 90\% CL on the numbers of $K^+\to\ell^+N$ events $N_S^\ell$ obtained from these inputs. (a) $K^+\to e^+N$ analysis; (b): $K^+\to\mu^+N$ analysis. For completeness, the squared mass scale is also shown. The legend shown in Figure (b) refers to both panels.}
\label{fig:ul-events}
\end{figure}

In each HNL mass hypothesis, the total number of observed events $N_{\rm obs}$ within the $\pm1.5\sigma_m^\ell$ HNL search window, the number of expected background events $N_{\rm exp}$ and its uncertainty $\delta N_{\rm exp}$ are used to compute confidence intervals for the number of observed $K^+\to\ell^+N$ decays $N_S^\ell$. The Rolke-L\'opez method~\cite{ro01} assuming Poissonian (Gaussian) distributions for the numbers of observed (expected) events is used. The procedure has been tested and found to be unbiased in the presence of artificially injected statistically significant $K^+\to\ell^+N$ signals. The values of $N_{\rm exp}$, $\delta N_{\rm exp}$ and $N_{\rm obs}$ in each HNL mass hypothesis considered are shown in Fig.~\ref{fig:ul-events}. The maximum value of the local signal significance computed as
\begin{displaymath}
z=(N_{\rm obs}-N_{\rm exp})/\sqrt{N_{\rm obs}+(\delta N_{\rm exp})^2}
\end{displaymath}
is 2.2, for the $e^+$ case with $m_N=283~{\rm MeV}/c^2$. In the absence of statistically significant HNL production signals, upper limits on $N_S^\ell$ are established; the expected and observed limits at 90\% CL are shown in Fig.~\ref{fig:ul-events}. Perfect knowledge of the background ($\delta N_{\rm exp}=0$) would improve these limits typically by 30\%.

\begin{figure}[p]
\begin{center}
\resizebox{0.50\textwidth}{!}{\includegraphics{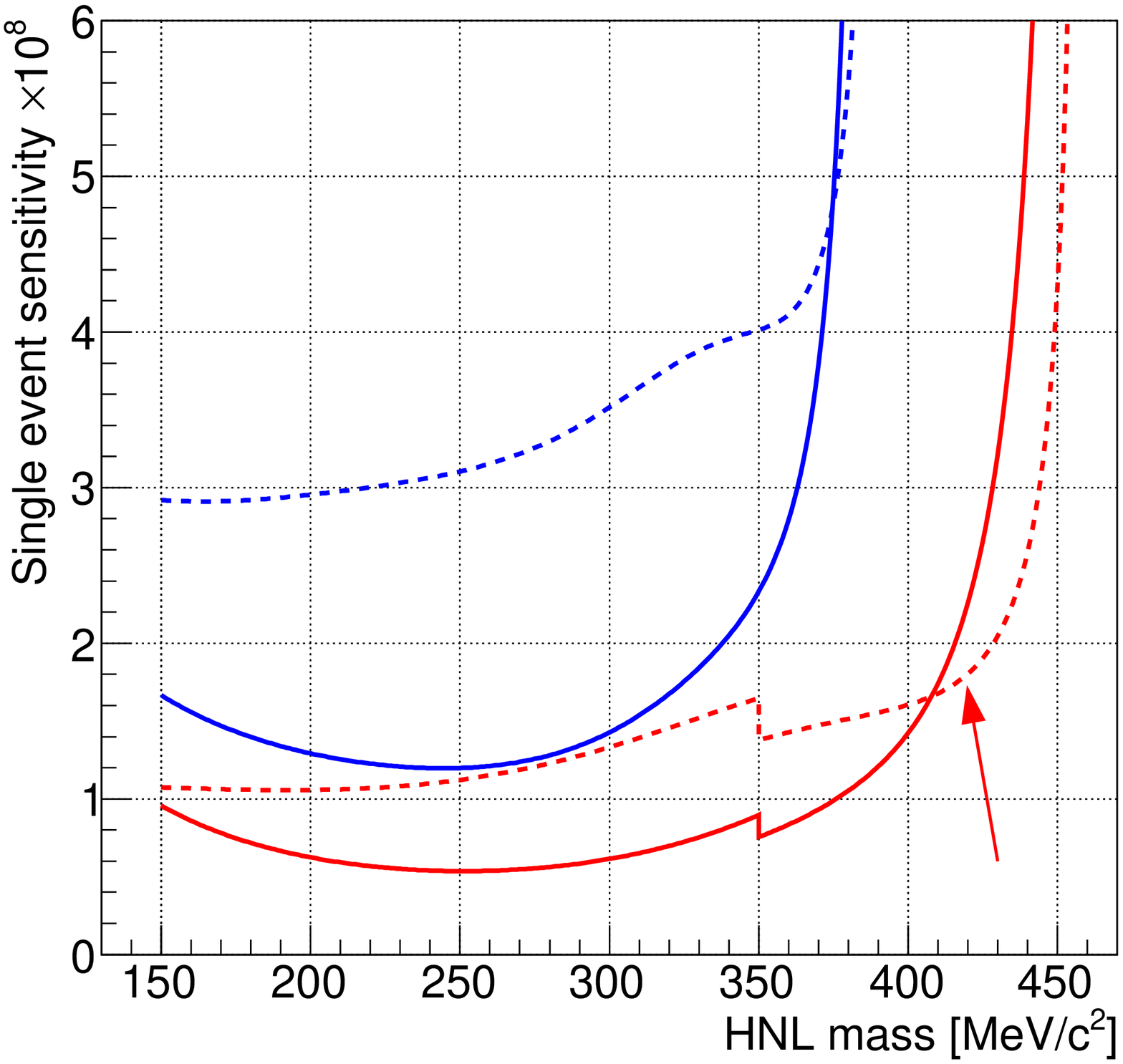}}%
\resizebox{0.50\textwidth}{!}{\includegraphics{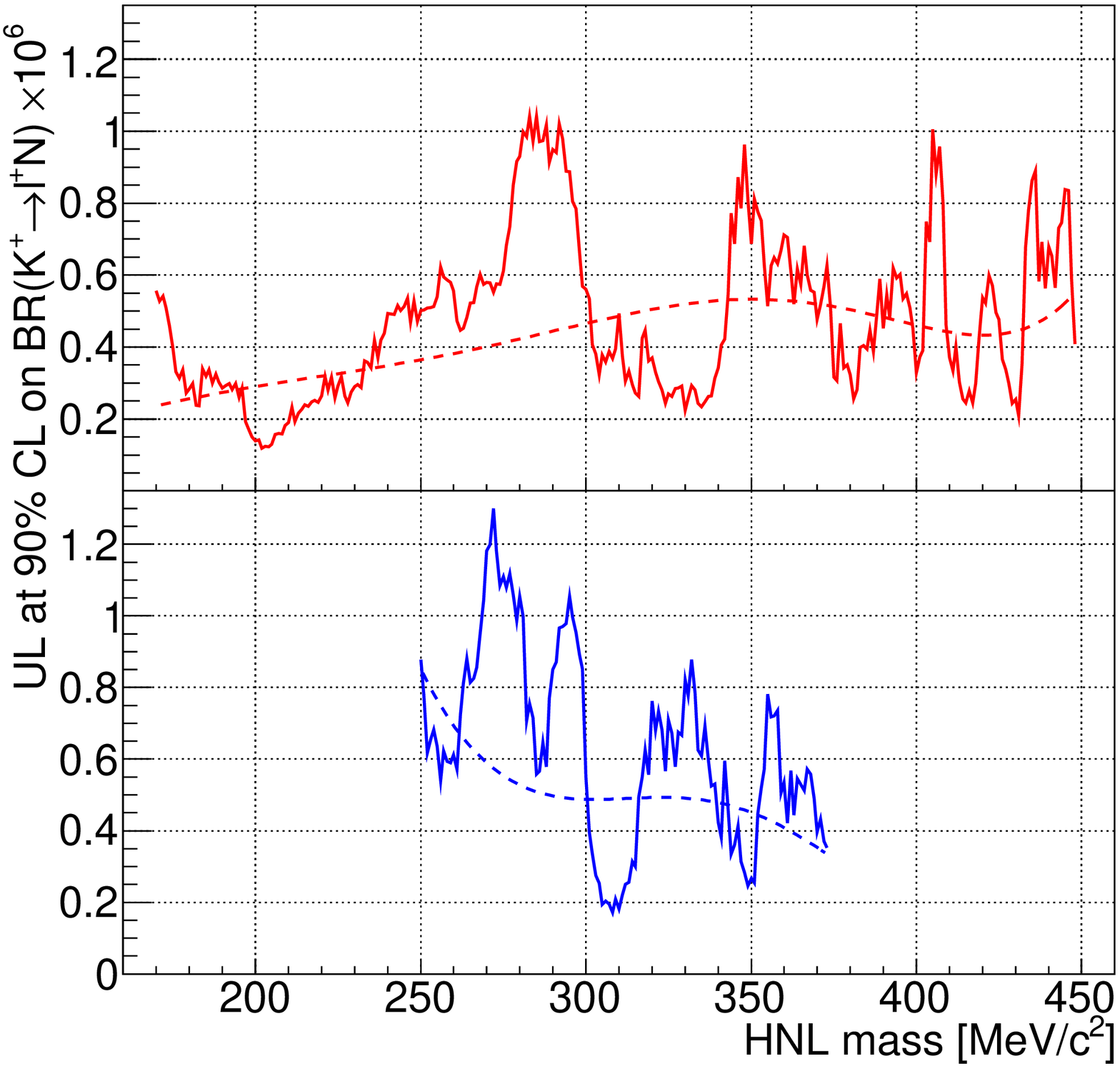}}%
\put(-430,200){\bf\large (a)} \put(-196,200){\bf\large (b)}
\boldmath
\put(-302, 33){\scriptsize\color{red}${\cal B}_{\rm SES}(K^+\!\!\to\!\! e^+\!N)$}
\put(-420,130){\scriptsize\color{blue}${\cal B}_{\rm SES}(K^+\!\!\to\!\!\mu^+\!N)$}
\put(-281,200){\tiny\color{red}$|U_{e4}|^2_{\rm SES}$}
\put(-340,100){\scriptsize\color{blue}$|U_{\mu4}|^2_{\rm SES}$}
\put(-71,205){\color{red}$K^+\!\to\! e^+\!N$}
\put(-71,107){\color{blue}$K^+\!\to\!\mu^+N$}
\unboldmath
\end{center}
\vspace{-13mm}
\caption{(a) Single event sensitivities ${\cal B}_{\rm SES}(K^+\to\ell^+N)$ (dashed lines) and $|U_{\ell 4}|^2_{\rm SES}$ (solid lines) defined in the text as functions of the assumed HNL mass. (b) Expected (dashed lines) and observed (solid lines) upper limits at 90\% CL on ${\cal B}(K^+\to e^+N)$ (top panel) and ${\cal B}(K^+\to\mu^+N)$ (bottom panel) obtained for each HNL mass hypothesis.}
\label{fig:ses}
\end{figure}

Single event sensitivities (SES) defined as the values of ${\cal B}(K^+\to\ell^+N)$ and the mixing parameter $|U_{\ell 4}|^2$ corresponding to the observation of one signal event,
\begin{displaymath}
{\cal B}_{\rm SES}(K^+\to\ell^+N) = \frac{1}{N_K^\ell \cdot A_\ell^N} ~~~~ {\rm and} ~~~~
|U_{\ell 4}|^2_{\rm SES} = \frac{{\cal B}_{\rm SES}(K^+\to\ell^+N)}{{\cal B}(K^+\to\ell^+\nu) \cdot \rho_\ell (m_N)},
\end{displaymath}
are displayed as functions of HNL mass in Fig.~\ref{fig:ses}a. They are ${\cal O}(10^{-8})$, and those in the positron case are smaller than those in the muon case due to $N_K^e$ being larger than $N_K^\mu$.

Upper limits on the branching fraction ${\cal B}(K^+\to\ell^+N)$ in each HNL mass hypothesis are computed from those on $N_S^\ell$ using eq.~(\ref{eq:master}); the expected and observed limits at 90\% CL are shown in Fig.~\ref{fig:ses}b. Upper limits on the mixing parameter $|U_{\ell 4}|^2$ in each HNL mass hypothesis are computed from those on ${\cal B}(K^+\to\ell^+N)$ according to eq.~(\ref{eq:main}). These limits depend on the external inputs ${\cal B}(K^+\to\ell^+\nu)$ only in the $e^+$ case due to the background subtraction in the $N_K^e$ computation. Systematic uncertainties on the limits are, in relative terms, of the same magnitude as those on $N_K^\ell$ (Section~\ref{sec:flux}).

\begin{figure}[t]
\begin{center}
\resizebox{0.60\textwidth}{!}{\includegraphics{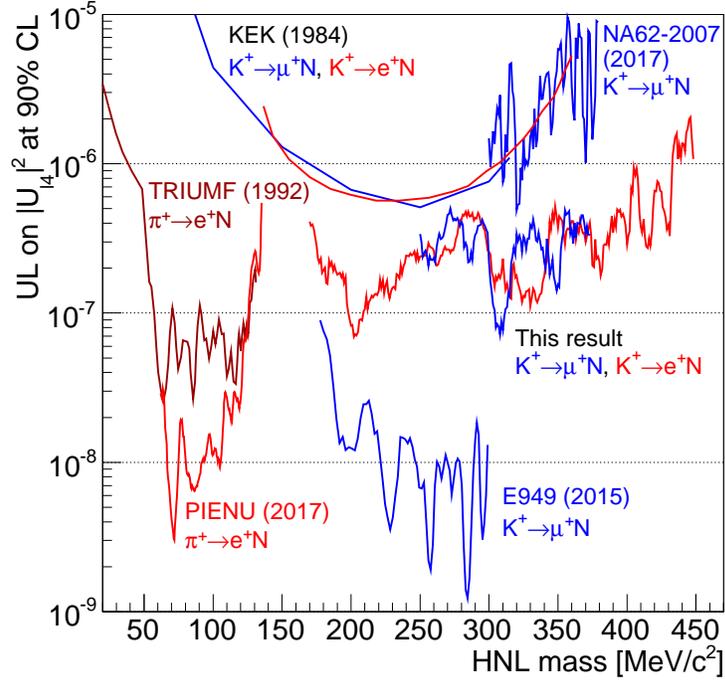}}
\end{center}
\vspace{-14mm}
\caption{Upper limits at 90\% CL on $|U_{\ell 4}|^2$ obtained for each assumed HNL mass compared to the limits established by earlier HNL production searches in $\pi^+$ decays:
TRIUMF (1992)~\cite{br92}, PIENU (2017)~\cite{ag17} and $K^+$ decays: KEK (1984)~\cite{ya84}, E949 (2015)~\cite{ar15}, NA62-2007 (2017)~\cite{la17}.}
\label{fig:world}
\end{figure}

The obtained upper limits on $|U_{\ell 4}|^2$ at 90\% CL together with the limits from previous HNL production searches in $\pi^+$~\cite{br92,ag17} and $K^+$~\cite{ya84,ar15,la17} decays are shown in Fig.~\ref{fig:world}. The reported result improves the existing limits on both $|U_{e4}|^2$ (over the whole mass range considered) and $|U_{\mu 4}|^2$ (above 300~MeV/$c^2$).

\section*{Summary}

A search for HNL production in $K^+\to\ell^+N$ decays has been performed with NA62 data  recorded in 2015 at $\sim 1\%$ of the nominal beam intensity with a minimum bias trigger. Upper limits on the HNL mixing parameters $|U_{e4}|^2$ and $|U_{\mu4}|^2$ in the ranges 170--448~MeV/$c^2$ and 250--373~MeV/$c^2$, respectively, have been established at the level between $10^{-7}$ and $10^{-6}$. This improves on the previous limits from HNL production searches over the whole mass range considered for $|U_{e4}|^2$ (and extends the mass range in which the limits exist), and above $m_N=300$~MeV/$c^2$ for $|U_{\mu4}|^2$.


\section*{Acknowledgements}

It is a pleasure to express our appreciation to the staff of the CERN laboratory and the technical staff of the participating laboratories and universities for their efforts in the operation of the experiment and data processing.

\input{acknow2015}



\end{document}

%% file: hnl2015_authors_6.tex
\begin{center}
{\Large The NA62 Collaboration$\,$\renewcommand{\thefootnote}{\fnsymbol{footnote}}%
\footnotemark[1]\renewcommand{\thefootnote}{\arabic{footnote}}}\\
\end{center}
\vspace{3mm}
\begin{raggedright}
\noindent
{\bf Universit\'e Catholique de Louvain, Louvain-La-Neuve, Belgium}\\
 E.~Cortina Gil,
 E.~Minucci,
 S.~Padolski$\,$\footnotemark[1],
 P.~Petrov,
 B.~Velghe$\,$\footnotemark[2]\\[2mm]
{\bf Faculty of Physics, University of Sofia, Sofia, Bulgaria}\\
 G.~Georgiev$\,$\footnotemark[3],
 V.~Kozhuharov$\,$\footnotemark[3],
 L.~Litov\\[2mm]
{\bf TRIUMF, Vancouver, British Columbia, Canada}\\
 T.~Numao\\[2mm]
{\bf University of British Columbia, Vancouver, British Columbia, Canada}\\
 D.~Bryman,
 J.~Fu$\,$\footnotemark[4]\\[2mm]
{\bf Charles University, Prague, Czech Republic}\\
 T.~Husek$\,$\footnotemark[5],
 K.~Kampf,
 M.~Zamkovsky\\[2mm]
{\bf Institut f\"ur Physik and PRISMA Cluster of excellence, Universit\"at Mainz, Mainz, Germany}\\
 R.~Aliberti,
 G.~Khoriauli,
 J.~Kunze,
 D.~Lomidze$\,$\footnotemark[6],
 R.~Marchevski,
 L.~Peruzzo,
 M.~Vormstein,
 R.~Wanke\\[2mm]
{\bf Dipartimento di Fisica e Scienze della Terra dell'Universit\`a e INFN, Sezione di Ferrara, Ferrara, Italy}\\
 P.~Dalpiaz,
 M.~Fiorini,
 E.~Gamberini$\,$\footnotemark[7],
 I.~Neri,
 A.~Norton,
 F.~Petrucci,
 H.~Wahl\\[2mm]
{\bf INFN, Sezione di Ferrara, Ferrara, Italy}\\
 A.~Cotta Ramusino,
 A.~Gianoli\\[2mm]
{\bf Dipartimento di Fisica e Astronomia dell'Universit\`a e INFN, Sezione di Firenze, Sesto Fiorentino, Italy}\\
 E.~Iacopini,
 G.~Latino,
 M.~Lenti\\[2mm]
{\bf INFN, Sezione di Firenze, Sesto Fiorentino, Italy}\\
 A.~Bizzeti$\,$\footnotemark[8],
 F.~Bucci,
 R.~Volpe\\[2mm]
{\bf Laboratori Nazionali di Frascati, Frascati, Italy}\\
 A.~Antonelli,
 F.~Gonnella$\,$\footnotemark[9],
 G.~Lamanna$\,$\footnotemark[10],
 G.~Lanfranchi,
 G.~Mannocchi,
 S.~Martellotti,
 M.~Moulson,
 M.~Raggi$\,$\footnotemark[11],
 T.~Spadaro\\[2mm]
{\bf Dipartimento di Fisica ``Ettore Pancini'' e INFN, Sezione di Napoli, Napoli, Italy}\\
 F.~Ambrosino,
 T.~Capussela,
 M.~Corvino,
 D.~Di Filippo,
 P.~Massarotti,
 M.~Mirra,
 M.~Napolitano,
 G.~Saracino\\[2mm]
{\bf Dipartimento di Fisica e Geologia dell'Universit\`a e INFN, Sezione di Perugia, Perugia, Italy}\\
 G.~Anzivino,
 F.~Brizioli,
 E.~Imbergamo,
 R.~Lollini,
 C.~Santoni\\[2mm]
{\bf INFN, Sezione di Perugia, Perugia, Italy}\\
 M.~Barbanera$\,$\footnotemark[12],
 P.~Cenci,
 B.~Checcucci,
 V.~Duk$\,$\footnotemark[9],
 P.~Lubrano,
 M.~Lupi$\,$\footnotemark[7],
 M.~Pepe,
 M.~Piccini\\[2mm]
{\bf Dipartimento di Fisica dell'Universit\`a e INFN, Sezione di Pisa, Pisa, Italy}\\
 F.~Costantini,
 L.~Di Lella,
 N.~Doble,
 M.~Giorgi,
 S.~Giudici,
 E.~Pedreschi,
 M.~Sozzi\\[2mm]
{\bf INFN, Sezione di Pisa, Pisa, Italy}\\
 C.~Cerri,
 R.~Fantechi,
 R.~Piandani,
 J.~Pinzino$\,$\footnotemark[7],
 L.~Pontisso,
 F.~Spinella\\[2mm]
{\bf Scuola Normale Superiore e INFN, Sezione di Pisa, Pisa, Italy}\\
 I.~Mannelli\\[2mm]
{\bf Dipartimento di Fisica, Sapienza Universit\`a di Roma e INFN, Sezione di Roma I, Roma, Italy}\\
 G.~D'Agostini\\[2mm]
{\bf INFN, Sezione di Roma I, Roma, Italy}\\
 A.~Biagioni,
 E.~Leonardi,
 A.~Lonardo,
 P.~Valente,
 P.~Vicini\\[2mm]
{\bf INFN, Sezione di Roma Tor Vergata, Roma, Italy}\\
 R.~Ammendola,
 V.~Bonaiuto$\,$\footnotemark[13],
 N.~De Simone$\,$\footnotemark[7],
 L.~Federici$\,$\footnotemark[7],
 A.~Fucci,
 A.~Salamon,
 F.~Sargeni$\,$\footnotemark[14]\\[2mm]
{\bf Dipartimento di Fisica dell'Universit\`a e INFN, Sezione di Torino, Torino, Italy}\\
 R.~Arcidiacono$\,$\footnotemark[15],
 B.~Bloch-Devaux,
 M.~Boretto,
 L.~Iacobuzio$\,$\footnotemark[9],
 E.~Menichetti,
 E.~Migliore,
 D.~Soldi\\[2mm]
{\bf INFN, Sezione di Torino, Torino, Italy}\\
 C.~Biino,
 A.~Filippi,
 F.~Marchetto\\[2mm]
{\bf Instituto de F\'isica, Universidad Aut\'onoma de San Luis Potos\'i, San Luis Potos\'i, Mexico}\\
 J.~Engelfried,
 N.~Estrada-Tristan\\[2mm]
{\bf Horia Hulubei national Institute of Physics and Nuclear Engineering, Bucharest-Magurele, Romania}\\
 A. M.~Bragadireanu,
 S. A.~Ghinescu,
 O. E.~Hutanu\\[2mm]
{\bf Joint Institute for Nuclear Research, Dubna, Russia}\\
 T.~Enik,
 V.~Falaleev,
 V.~Kekelidze,
 A.~Korotkova,
 D.~Madigozhin,
 M.~Misheva$\,$\footnotemark[16],
 N.~Molokanova,
 S.~Movchan,
 I.~Polenkevich,
 Yu.~Potrebenikov,
 S.~Shkarovskiy,
 A.~Zinchenko$\,$\renewcommand{\thefootnote}{\fnsymbol{footnote}}\footnotemark[2]\renewcommand{\thefootnote}{\arabic{footnote}}\\[2mm]
{\bf Institute for Nuclear Research of the Russian Academy of Sciences, Moscow, Russia}\\
 S.~Fedotov,
 E.~Gushchin,
 A.~Khotyantsev,
 A.~Kleimenova$\,$\footnotemark[17],
 Y.~Kudenko$\,$\footnotemark[18],
 V.~Kurochka,
 M.~Medvedeva,
 A.~Mefodev,
 A.~Shaikhiev\\[2mm]
{\bf Institute for High Energy Physics - State Research Center of Russian Federation, Protvino, Russia}\\
 S.~Kholodenko,
 V.~Kurshetsov,
 V.~Obraztsov,
 A.~Ostankov,
 V.~Semenov,
 V.~Sugonyaev,
 O.~Yushchenko\\[2mm]
{\bf Faculty of Mathematics, Physics and Informatics, Comenius University, Bratislava, Slovakia}\\
 L.~Bician,
 T.~Blazek,
 V.~Cerny,
 M.~Koval$\,$\footnotemark[7],
 Z.~Kucerova\\[2mm]
{\bf CERN,  European Organization for Nuclear Research, Geneva, Switzerland}\\
 A.~Ceccucci,
 H.~Danielsson,
 F.~Duval,
 B.~D\"obrich,
 L.~Gatignon,
 R.~Guida,
 F.~Hahn,
 B.~Jenninger,
 P.~Laycock,
 G.~Lehmann Miotto,
 P.~Lichard,
 A.~Mapelli,
 M.~Noy,
 V.~Palladino$\,$\footnotemark[19],
 M.~Perrin-Terrin$\,$\footnotemark[17],
 G.~Ruggiero$\,$\footnotemark[20],
 V.~Ryjov,
 S.~Venditti\\[2mm]
{\bf University of Birmingham, Birmingham, United Kingdom}\\
 M. B.~Brunetti,
 V.~Fascianelli$\,$\footnotemark[21],
 E.~Goudzovski,
 C.~Lazzeroni,
 N.~Lurkin,
 F.~Newson,
 C.~Parkinson,
 A.~Romano,
 A.~Sergi,
 A.~Sturgess,
 J.~Swallow\\[2mm]
{\bf University of Bristol, Bristol, United Kingdom}\\
 H.~Heath,
 R.~Page,
 S.~Trilov\\[2mm]
{\bf University of Glasgow, Glasgow, United Kingdom}\\
 B.~Angelucci,
 D.~Britton,
 C.~Graham,
 D.~Protopopescu\\[2mm]
{\bf University of Liverpool, Liverpool, United Kingdom}\\
 J. B.~Dainton$\,$\footnotemark[20],
 J. R.~Fry$\,$\footnotemark[9],
 L.~Fulton,
 D.~Hutchcroft,
 K.~Massri$\,$\footnotemark[7],
 E.~Maurice$\,$\footnotemark[22],
 B.~Wrona\\[2mm]
{\bf George Mason University, Fairfax, Virginia, USA}\\
 A.~Conovaloff,
 P.~Cooper,
 D.~Coward$\,$\footnotemark[23],
 P.~Rubin\\[2mm]
\end{raggedright}
%
%
\setcounter{footnote}{0}
\renewcommand{\thefootnote}{\fnsymbol{footnote}}
\footnotetext[1]{Corresponding author: E.~Goudzovski, email: Evgueni.Goudzovski@cern.ch}
\footnotetext[2]{Deceased}
\renewcommand{\thefootnote}{\arabic{footnote}}
\footnotetext[1]{Present address: Brookhaven National Laboratory, Upton, NY 11973, USA}
\footnotetext[2]{Present address: TRIUMF, Vancouver, British Columbia, V6T 2A3, Canada}
\footnotetext[3]{Also at Laboratori Nazionali di Frascati, I-00044 Frascati, Italy}
\footnotetext[4]{Present address: UCLA Physics and Biology in Medicine, Los Angeles, CA 90095, USA}
\footnotetext[5]{Also at School of Physics and Astronomy, University of Birmingham, Birmingham, B15 2TT, UK}
\footnotetext[6]{Present address: Universit\"at Hamburg, D-20146 Hamburg, Germany}
\footnotetext[7]{Present address: CERN,  European Organization for Nuclear Research, CH-1211 Geneva 23, Switzerland}
\footnotetext[8]{Also at Dipartimento di Fisica, Universit\`a di Modena e Reggio Emilia, I-41125 Modena, Italy}
\footnotetext[9]{Present address: School of Physics and Astronomy, University of Birmingham, Birmingham, B15 2TT, UK}
\footnotetext[10]{Present address: Dipartimento di Fisica dell'Universit\`a e INFN, Sezione di Pisa, I-56100 Pisa, Italy}
\footnotetext[11]{Present address: Dipartimento di Fisica, Universit\`a di Roma La Sapienza, I-00185 Roma, Italy}
\footnotetext[12]{Present address: INFN, Sezione di Pisa, I-56100 Pisa, Italy}
\footnotetext[13]{Also at Department of Industrial Engineering, University of Roma Tor Vergata, I-00173 Roma, Italy}
\footnotetext[14]{Also at Department of Electronic Engineering, University of Roma Tor Vergata, I-00173 Roma, Italy}
\footnotetext[15]{Also at Universit\`a degli Studi del Piemonte Orientale, I-13100 Vercelli, Italy}
\footnotetext[16]{Present address: Institute of Nuclear Research and Nuclear Energy of Bulgarian Academy of Science (INRNE-BAS), BG-1784 Sofia, Bulgaria}
\footnotetext[17]{Present address: Universit\'e Catholique de Louvain, B-1348 Louvain-La-Neuve, Belgium}
\footnotetext[18]{Also at National Research Nuclear University (MEPhI), 115409 Moscow and Moscow Institute of Physics and Technology, 141701 Moscow region, Moscow, Russia}
\footnotetext[19]{Present address: Physics Department, Imperial College London, London, SW7 2BW, UK}
\footnotetext[20]{Present address: Physics Department, University of Lancaster, Lancaster, LA1 4YW, UK}
\footnotetext[21]{Present address: Dipartimento di Psicologia, Universit\`a di Roma La Sapienza, I-00185 Roma, Italy}
\footnotetext[22]{Present address: Laboratoire Leprince Ringuet, F-91120 Palaiseau, France}
\footnotetext[23]{Also at SLAC National Accelerator Laboratory, Stanford University, Menlo Park, CA 94025, USA}

%% file: acknow2015.tex
The cost of the experiment and of its auxiliary systems were supported by the funding agencies of
the Collaboration Institutes. We are particularly indebted to:
F.R.S.-FNRS (Fonds de la Recherche Scientifique - FNRS), Belgium;
BMES (Ministry of Education, Youth and Science), Bulgaria;
NSERC (Natural Sciences and Engineering Research Council), Canada;
NRC (National Research Council) contribution to TRIUMF, Canada;
MEYS (Ministry of Education, Youth and Sports),  Czech Republic;
BMBF (Bundesministerium f\"{u}r Bildung und Forschung) contracts 05H12UM5 and 05H15UMCNA, Germany;
INFN  (Istituto Nazionale di Fisica Nucleare), Italy;
MIUR (Ministero dell'Istruzione, dell'Universit\`a e della Ricerca),  Italy;
CONACyT  (Consejo Nacional de Ciencia y Tecnolog\'{i}a),  Mexico;
IFA (Institute of Atomic Physics),  Romania;
INR-RAS (Institute for Nuclear Research of the Russian Academy of Sciences), Moscow, Russia;
JINR (Joint Institute for Nuclear Research), Dubna, Russia;
NRC (National Research Center)  ``Kurchatov Institute'' and MESRF (Ministry of Education and Science of the Russian Federation), Russia;
MESRS  (Ministry of Education, Science, Research and Sport), Slovakia;
CERN (European Organization for Nuclear Research), Switzerland;
STFC (Science and Technology Facilities Council), United Kingdom;
NSF (National Science Foundation) Award Number 1506088,   U.S.A.;
ERC (European Research Council)  ``UniversaLepto" advanced grant 268062, ``KaonLepton" starting grant 336581, Europe.

Individuals have received support from:
Charles University (project GA UK number 404716), Czech Republic;
Ministry of Education, Universities and Research (MIUR  ``Futuro in ricerca 2012''  grant RBFR12JF2Z, Project GAP), Italy;
the Royal Society (grants UF100308, UF0758946), United Kingdom;
STFC (Rutherford fellowships ST/J00412X/1, ST/M005798/1), United Kingdom;
ERC (grants 268062,  336581).